\documentclass[conference]{IEEEtran}
\IEEEoverridecommandlockouts
\usepackage{cite}
\usepackage{stfloats}
\usepackage{url}
\usepackage{amsmath,amssymb,amsfonts}
\usepackage{algorithmic}
\usepackage{graphicx}
\usepackage{textcomp}
\usepackage{xcolor}
\usepackage{booktabs}
\usepackage{colortbl}
\usepackage{array}
\def\BibTeX{{\rm B\kern-.05em{\sc i\kern-.025em b}\kern-.08em
    T\kern-.1667em\lower.7ex\hbox{E}\kern-.125emX}}
\begin{document}

\title{A Practical Two-Stage Framework for GPU Resource and Power Prediction in Heterogeneous HPC Systems
}

\makeatletter
\newcommand{\linebreakand}{%
\end{@IEEEauthorhalign}
\hfill\mbox{}\par
\mbox{}\hfill\begin{@IEEEauthorhalign}
}
\makeatother

\author{

\IEEEauthorblockN{Beste Oztop}
\IEEEauthorblockA{\textit{Electrical and Computer Engineering}
\\
\textit{Boston University} \\
Boston, MA \\
boztop@bu.edu} 
\and
\IEEEauthorblockN{Dhruva Kulkarni}
\IEEEauthorblockA{\textit{Application Performance Group (NERSC)} \\
\textit{Lawrence Berkeley National Laboratory} \\
Berkeley, CA \\
dkulkarni@lbl.gov} 
\and
\IEEEauthorblockN{Zhengji Zhao}
\IEEEauthorblockA{\textit{Advanced Technologies Group (NERSC)} \\
\textit{Lawrence Berkeley National Laboratory} \\
Berkeley, CA \\
zzhao@lbl.gov} 

\linebreakand
\IEEEauthorblockN{Ayse K. Coskun}
\IEEEauthorblockA{\textit{Electrical and Computer Engineering}
\\
\textit{Boston University} \\
Boston, MA \\
acoskun@bu.edu} 

\and
\IEEEauthorblockN{Kadidia Konate}
\IEEEauthorblockA{\textit{Data \& AI Sciences Group (NERSC)} \\
\textit{Lawrence Berkeley National Laboratory} \\
Berkeley, CA \\
kadidiakonate@lbl.gov} 

}

\maketitle

\begin{abstract}
Efficient utilization of GPU resources and power has become critical with the growing demand for GPUs in high-performance computing (HPC). In this paper, we analyze GPU utilization and GPU memory utilization, as well as the power consumption of the Vienna ab initio Simulation Package (VASP), using the Slurm workload manager historical logs and GPU performance metrics collected by NVIDIA’s Data Center GPU Manager (DCGM). VASP is a widely used materials science application on Perlmutter at NERSC, an HPE Cray EX system based on NVIDIA A100 GPUs. Using our insights from the resource utilization analysis of VASP applications, we propose a resource prediction framework to predict the average GPU power, maximum GPU utilization, and maximum GPU memory utilization values of heterogeneous HPC system applications to enable more efficient scheduling decisions and power-aware system operation. Our prediction framework consists of two stages: 1) using only the Slurm accounting logs as training data and 2) augmenting the training data with historical GPU profiling metrics collected with DCGM. The maximum GPU utilization predictions using only the Slurm submission features achieve up to 97\% accuracy. Furthermore, features engineered from GPU-compute and memory activity metrics exhibit good correlations with average power utilization, and our runtime power usage prediction experiments result in up to 92\% prediction accuracy. These findings demonstrate the effectiveness of DCGM metrics in capturing application characteristics and highlight their potential for developing predictive models to support dynamic power management in HPC systems.
\end{abstract}

\begin{IEEEkeywords}
High-Performance Computing, GPU Accelerated Applications, Resource Utilization Analysis, Resource Prediction
\end{IEEEkeywords}

\section{Introduction}\label{sec1}


High-Performance Computing (HPC) systems support applications and simulations from various scientific domains, such as climate modeling, genetics, bioinformatics, and materials science, which continue to grow in input size and execution time as we reach the exascale era. As a result, there is an increasing demand for accelerators such as GPUs to process larger workloads more efficiently than traditional CPU-only nodes. However, due to the unpredictable nature of GPU node requirements for different workloads, there is often a mismatch between the resources users request and those they utilize, and memory utilization on GPU nodes is frequently less than desired \cite{li2023analyzing}. Thus, it is imperative to analyze and characterize the resource utilization of heterogeneous HPC applications. 

From a systems point of view, accurately predicting the power usage of GPU jobs, along with their GPU node and memory requirements, is important for efficient resource management, job scheduling, and energy optimization in HPC environments. Moreover, power-aware scheduling of jobs running on GPU nodes can reduce energy costs and improve overall system throughput, especially as GPUs become increasingly power-hungry components in modern supercomputers. Integrating resource provisioning frameworks with resource managers on HPC systems enables more sustainable and efficient HPC operations, especially in an era of growing environmental awareness and strict data center power budgets.

Motivated by these challenges, we focus on the analysis of resource utilization of the Vienna ab initio Simulation Package (VASP) \cite{hafner2008ab} applications submitted to the GPU nodes of the National Energy Research Scientific Computing Center (NERSC) Perlmutter system \cite{perlmutter_docs} in March 2025. Analyzing this 1-month-long data, comprising 32,322 VASP jobs, reveals the characteristics of resource utilization and power usage trends of Perlmutter's top GPU workload. Building on this data analysis, we present a two-stage framework for predicting power and resource usage for GPU jobs across the Perlmutter system. Our methodology generalizes to other GPU-accelerated applications and other large-scale GPU systems as long as similar GPU activity metrics are available\footnote{For reproducibility, we open-source the resource utilization analysis notebooks and prediction framework scripts at: \url{https://github.com/peaclab/VASP_gpu_jobs_analysis}.}. The contributions of this work are as follows:
\begin{itemize}
\item We present a large-scale data analysis of Slurm accounting data and NVIDIA DCGM metrics collected from VASP GPU jobs on the Perlmutter system, characterizing the average power consumption, maximum GPU utilization, and maximum GPU memory utilization in a production HPC environment.

\item Based on our analysis of the resource utilization data, we propose a practical, low-overhead, ML-based two-stage prediction framework to support system-level resource management and power-aware operation.
\begin{itemize}
        \item We estimate the total average power usage and maximum resource utilization using only Slurm submission features before jobs are submitted, enabling informed resource selection and improved scheduling decisions.
        \item We predict the average power consumption of GPU jobs at runtime using time-series DCGM GPU monitoring metrics, demonstrating a low-overhead and application-aware dynamic power management methodology.
    \end{itemize}
\end{itemize}

The remainder of this paper is organized as follows: We provide an overview of the related work in Section~\ref{sec2} and describe the system configuration and dataset characteristics in Section~\ref{sec3}. Section~\ref{sec4} presents the data analysis of the resource and power utilization of VASP GPU jobs. We introduce our prediction framework in Section~\ref{sec5} and present our experimental results in Section~\ref{sec6}. Section~\ref{disc} supports the generalizability of our prediction framework, and Section~\ref{sec8} concludes with a discussion of key findings and future work directions.

\section{Related Work}\label{sec2}

Understanding and improving resource utilization in HPC begins with predictive analysis of job characteristics. Traditional approaches leverage historical Slurm logs \cite{slurmurl} to forecast resource needs or execution time before scheduling \cite{tanash2021ensemble,thonglek2019improving}. However, such static models provide limited insight into the dynamic GPU usage patterns of workloads like VASP. Complementary runtime analysis focuses on operational inefficiencies, using metrics from tools such as LDMS \cite{nichols2022resource,agelastos2014lightweight} to capture performance variations. Recent studies have revealed widespread GPU memory underutilization \cite{li2023analyzing}, explored co-scheduling opportunities for more efficient resource usage \cite{maloney2024analyzing}, and demonstrated the value of fine-grained telemetry (e.g., DCGM) for identifying resource imbalances \cite{sencan2025gpu}. 

A parallel research direction explores coarse-grained power modeling for system-wide energy management, estimating average or peak job power to support aggregate resource planning \cite{menear2025classification}. Karimi et al. \cite{karimi2024profiling} profiled power usage on a leadership-scale HPC system, revealing large variability while focusing only on aggregated trends. Antici et al. \cite{antici25isc} proposed an online CPU-centric power prediction framework. Similar studies on Summit \cite{summit2020temporal} and Polaris \cite{polaris2025energy} examined cluster-level power, offering limited insight into intra-job dynamics. Such high-level approaches remain inadequate for capturing fine-grained power fluctuations within running jobs, an essential capability for runtime energy optimization.

To capture these dynamics, recent work has advanced fine-grained runtime power prediction frameworks using high-resolution telemetry. These enable application-aware power management by dynamically allocating energy where it yields the most performance benefits. The TARDIS scheduler, for instance, employs a Graph Neural Network (GNN) for job-level power prediction to minimize operational costs \cite{tardis2025power}. Other methods adopt adaptive incremental learning for online prediction \cite{onlineincremental2023}, leverage architectural performance counters to forecast unseen workloads \cite{framework2024performance}, or extend early GPU power estimation research \cite{hpca2015gpubuild}. Their reliance on low-level performance data makes these models broadly applicable across heterogeneous GPU workloads.

Despite these advances, generalizable runtime prediction frameworks remain undervalidated on irregular scientific workloads in production HPC environments. Our work addresses this gap using VASP, which accounts for over 20\% of NERSC’s workload and exhibits complex, rapidly varying GPU power profiles. We introduce and validate a prediction pipeline built on NVIDIA’s DCGM metrics, a lightweight, live-access telemetry well-suited for runtime decision-making. This approach captures fast-changing GPU behavior, which is critical for power-aware optimization. We show that integrating DCGM time-series data significantly improves prediction accuracy (achieving between 81\% and 92\%) for dynamic GPU power compared to naive prediction methods. By demonstrating the approach with VASP, we establish a path toward generalizable, real-time power steering for diverse GPU-accelerated HPC applications.

\section{Methodology}\label{sec3}
\subsection{Perlmutter Architecture}
In this work, we use the application data from the NERSC Perlmutter system from March 2025. Perlmutter is an HPE (Hewlett-Packard Enterprise) Cray EX supercomputer that contains 3,072 CPU-only nodes and 1,792 GPU-accelerated nodes. We are mainly interested in the GPU nodes, each equipped with one AMD EPYC 7763 CPU, four NVIDIA A100 GPUs, and four HPE Slingshot 11 NICs. The A100 GPU memory is 40 GB on 1,536 nodes and 80 GB on the remaining 256 nodes. This work includes utilization data from all available GPU nodes across regular and premium Quality of Service (QoS) classes. Since the shared GPU nodes allow multi-tenancy and resource sharing of GPU jobs, we leave a more comprehensive resource utilization analysis of the shared nodes for future work.

\subsection{VASP} 
The Vienna Ab initio Simulation Package (VASP)~\cite{vasp} is a widely used quantum-mechanical molecular dynamics code, highly ranked at NERSC and other supercomputing centers in terms of the number of batch job submissions. It solves a non-linear eigenvalue problem iteratively through self-consistent cycles, forming the basis for its broad feature set and diverse computational workflows. VASP is primarily written in Fortran 90 and relies heavily on FFTs and linear algebra libraries (BLAS/LAPACK/ScaLAPACK). 

\subsection{Dataset Characteristics}
This work analyzes one-month (March 2025) of VASP jobs on Perlmutter using Slurm's accounting logs and GPU profiling metrics collected by NVIDIA’s Data Center GPU Manager (DCGM) \cite{nvidia_dcgm_docs}. VASP jobs account for at least 20.2\% of the total GPU jobs submitted to Perlmutter during this period, ranking first. The dataset with available DCGM metrics included in this work contains information on a subset of these workloads, comprising a total of 32,322 VASP jobs and over 5,480 GPU node-hours. Table~\ref{tab:slurm_selected_metrics} lists selected submission features from the Slurm workload manager. Note that in the Perlmutter system, the requested number of nodes directly determines the requested number of processors, the maximum memory size, and the number of GPUs requested, or the other way around. Other information from the job submission scripts, such as job input files, the working directory, and necessary modules, is not available in this study but could contribute to our findings in the future if it becomes available. 

\begin{table}[t]
\caption{Selected Slurm job accounting fields.}
\label{tab:slurm_selected_metrics}
\centering
\begin{tabular}{p{1.75cm}p{6cm}}
\hline
\textbf{Feature} & \textbf{Description} \\
\hline
User      & Name of the user submitting the job. \\
JobName   & User selected job name. \\
Account   & User's account to charge for computing. \\
Category  & Scientific category the user belongs to. \\
ReqCPUs   & Number of CPUs requested. \\
ReqNode   & Number of GPU nodes requested. \\
ReqGPUs   & Number of GPUs requested. \\
ReqMem    & Requested memory size. \\
Timelimit & Hard wallclock time limit. \\
\hline
\end{tabular}
\end{table}

GPU profiling data from DCGM is collected system-wide on Perlmutter via LDMS \cite{agelastos2014lightweight} at 10-second sampling intervals. Thus, unlike Slurm accounting fields, DCGM metrics are time series data that are collected for each GPU job throughout its execution. The DCGM metric collection from March 2025 that we include in this work contains some unintended data points, which appear at intervals of less than 10 seconds. These irregular data points can result from metric collection configurations or other debugging activities. For consistency, we filter these irregular data points out. Therefore, after the filtering operation, for any given job with $t_{exec}$ total execution time in seconds, there exist $\frac{t_{exec}}{10}$ different entries for a given DCGM metric per GPU allocated to the job. For our analysis, we focus on a subset of DCGM metrics, as described in Table~\ref{tab:dcgm_metrics}. Note that SM stands for streaming multiprocessor, and the ratio of cycles ranges from 0 to 1.

\begin{table}[t]
\caption{Selected GPU profiling metrics from NVIDIA DCGM.}
\label{tab:dcgm_metrics}
\centering
\begin{tabular}{p{2.6cm}p{5cm}}
\hline
\textbf{Metric} & \textbf{Description} \\
\hline
\texttt{gpu\_utilization} & GPU utilization activity (\%). \\
\texttt{fb\_used} & Used frame buffer memory (MB). \\
\texttt{fb\_free} & Free frame buffer memory (MB). \\
\texttt{sm\_active} & The ratio of cycles where an SM has at least 1 active warp. \\
\texttt{sm\_occupancy} & The ratio of the number of warps resident on an SM, relative to the theoretical maximum number of active warps. \\
\texttt{dram\_active} &  The ratio of cycles with active device memory; sending or receiving data.\\
\texttt{fp64\_active} & The ratio of cycles where FP64 pipe is active. \\
\texttt{tensor\_active} & The ratio of cycles where any tensor pipeline is active.\\
\texttt{power\_usage} & Average device power usage (W). \\
\hline
\end{tabular}
\end{table}

In addition to the available Slurm and DCGM data, we utilize the following GPU memory utilization metric, which we later refer to as \texttt{mem\_utilization}. The GPU memory utilization percentage indicates the ratio of utilized GPU memory resources to the total available \texttt{framebuffer} memory.

\begin{equation}
    \texttt{mem\_utilization(\%)}= \frac{\texttt{fb\_used}}{\texttt{fb\_used} + \texttt{fb\_free}}.
    \label{eqn:mem_util}
\end{equation}

Note that we match Slurm and DCGM metrics for each GPU job based on job ID, node list, and the job start and end times, as DCGM metrics are reported with timestamps and associated job and node IDs. We further identify VASP data based on the executable names recorded by Slurm. 
Using the DCGM metrics introduced in this section, we analyze the resource utilization and power consumption behavior of VASP GPU applications next.

\section{Resource and Power Utilization Analysis}\label{sec4}
In this section, we analyze the maximum GPU and GPU memory utilization, as well as the average power consumption distributions of March 2025 VASP GPU jobs. Our observations in this section can help users identify application bottlenecks by detecting underutilized or overburdened resources. Additionally, understanding resource utilization trends can inform future system procurement by showing when and where more resources will be required.

Using the Slurm accounting fields in Table~\ref{tab:slurm_selected_metrics}, we observe 128 unique Perlmutter users across 15 different scientific categories who submitted the VASP jobs included in our analysis. The execution time of VASP jobs included in this analysis reaches up to 24 hours, while the majority of jobs are executed in less than one hour. 

\begin{figure*}
\centering
    \includegraphics[width=0.9\textwidth]{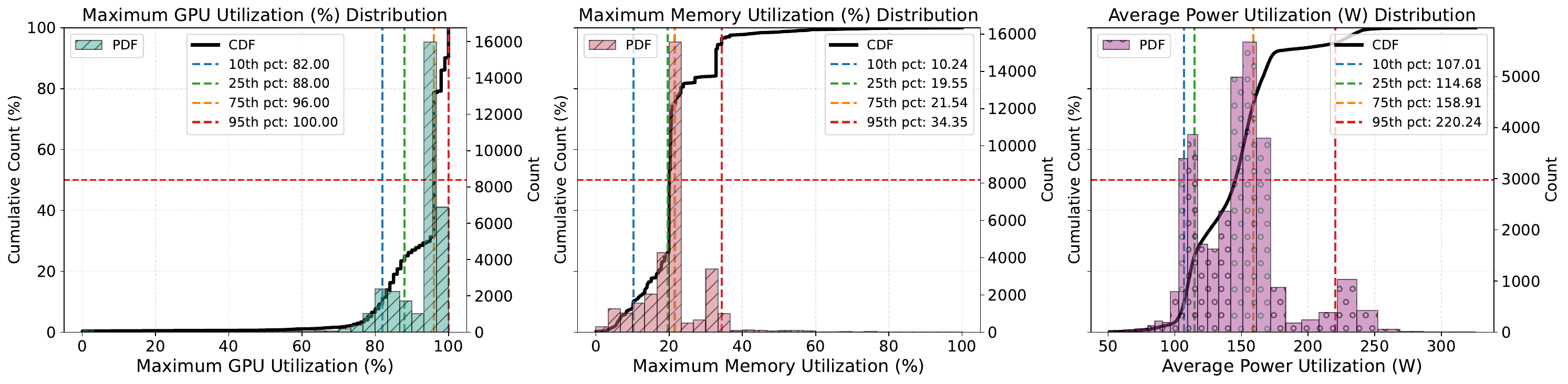}
    \caption{
    Distributions of maximum GPU utilization, maximum memory utilization, and average power for VASP GPU jobs on Perlmutter (March 2025).}
    \label{fig:res-util-dist}
    \vspace{-0.1in}
\end{figure*} 

Figure~\ref{fig:res-util-dist} presents the probability and cumulative density plots for maximum GPU utilization, GPU memory utilization, and average power usage, using the NVIDIA DCGM metrics presented in Table~\ref{tab:dcgm_metrics}. The percentage of GPU utilization is generally high, with the 10\textsuperscript{th} percentile at 82\% and the 95\textsuperscript{th} percentile reaching 100\%, indicating that at least 90\% of the VASP jobs utilize over 82\% of the GPU's capacity and that the majority of the jobs run at almost maximum GPU utilization. On the other hand, maximum GPU memory utilization is considerably lower and demonstrates more variance than maximum GPU utilization. The 10\textsuperscript{th} percentile is only 10.24\%, and even the 75\textsuperscript{th} percentile is at 21.54\%. This indicates that most VASP jobs utilize a small fraction of the available GPU memory. Only the top 5\% of jobs exceed 34.35\% GPU memory usage. This comparison highlights that the VASP jobs included in our analysis relatively underutilize the memory resources on GPU nodes. The average power usage of VASP jobs, on the other hand, exhibits a wide distribution. The 10\textsuperscript{th} percentile is at 107.01 W, meaning that 90\% of the VASP jobs have an average power utilization greater than 107.01 W. A total of 5\% of the jobs have an average power consumption of more than 220.24 W. While most VASP jobs consume moderate power, a small subset can be extremely power-intensive. 

These observations highlight the need to accurately predict GPU and GPU memory utilization for VASP jobs prior to execution on Perlmutter nodes. In addition, the fluctuations in their average power consumption across runs motivate the need to estimate average power usage both before and during job execution to enable more sustainable system operation. Based on these insights, we next present the design of our two-stage resource and power prediction framework.

\section{Framework Design}\label{sec5}

Our data analysis from the previous section reveals that the resource utilization characteristics of VASP jobs span a broad range of values. Predicting how much of the available resources a GPU job will utilize can help with better scheduling in heterogeneous HPC systems, as we can match workloads to nodes based on their resource usage profiles. Moreover, application-aware power management can further increase the energy efficiency of large-scale computing systems. Therefore, in this section, we present a two-stage framework to predict the resource utilization of VASP jobs and their average power utilization, which can be generalized to other applications. Figure~\ref{fig:overview-framework} provides an overview of our framework, which 1) predicts resource utilization values before jobs start executing, using Slurm features only, and 2) integrates DCGM runtime metrics for time-series-based average power prediction, which can enable dynamic power management practices.  

\begin{figure*}
\centering
    \includegraphics[width=0.8\textwidth]{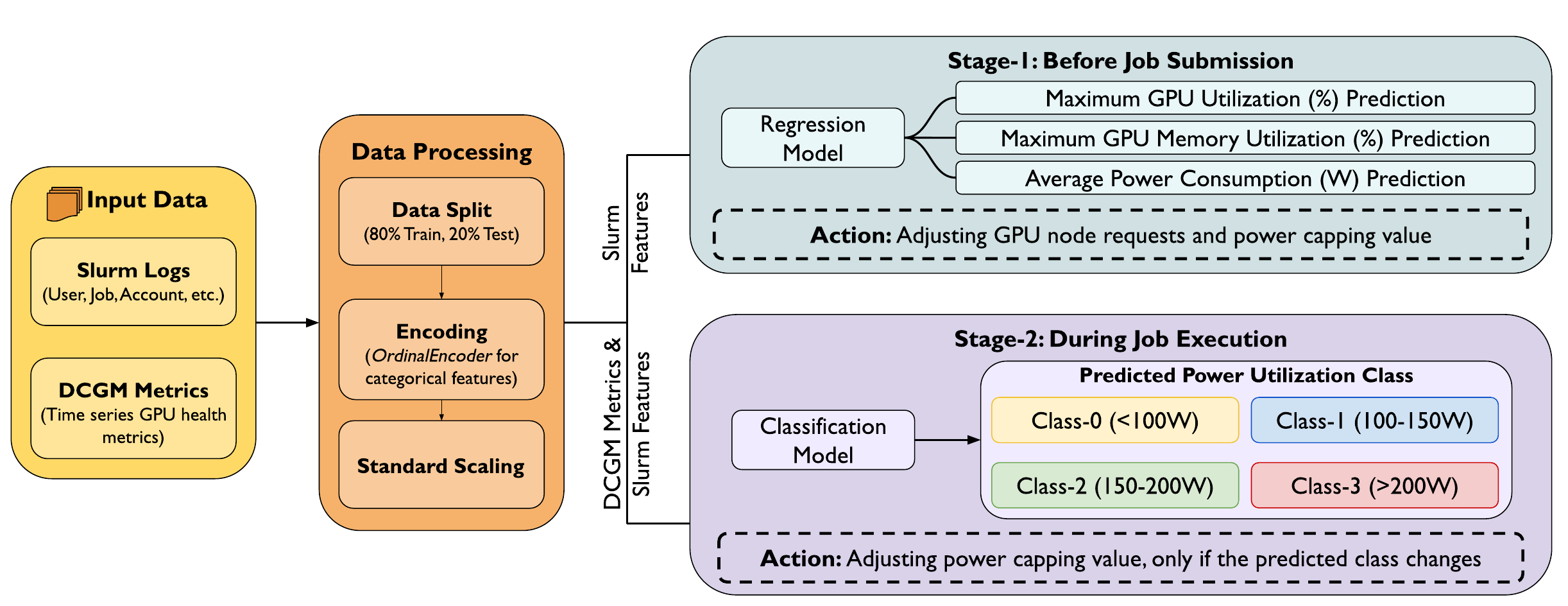}
    \caption{Overview of the proposed two-stage framework for GPU resource and power prediction.}
    \label{fig:overview-framework}
\end{figure*} 

\paragraph{Target Resource Utilization Prediction Variables} Following our discussion in the previous section, our framework predicts GPU utilization, GPU memory utilization, and average GPU power consumption, as these constitute a set of features that are crucial for understanding the characteristics of a GPU application. The target variables are aggregated metrics for both stages of our framework. For example, the maximum GPU and memory utilization percentages are the maximum values of all reported \texttt{gpu\_utilization} and calculated \texttt{mem\_utilization} metrics. These two metrics reflect how effectively an application leverages the computational power and physical GPU memory of its assigned GPU nodes. Therefore, predicting them before job submission provides valuable insights for optimizing resource allocation. Specifically, it enables users to select an appropriate number of GPU nodes before job execution, consequently improving system efficiency and reducing resource waste. Average power utilization prediction in both stages of the framework effectively reflects overall energy consumption (the product of power and runtime) and remains a robust alternative when instantaneous power prediction is not possible. This metric captures energy usage patterns and can be leveraged for tasks such as power capping. Note that the average power is calculated as the mean of all \texttt{avg\_power} values reported by the DCGM tool at 10-second intervals across all GPUs assigned to the job.

\paragraph{Motivation and Design of the ML-Based Prediction Framework} We employ machine learning (ML) models in our framework because the resource and power utilization data from VASP GPU jobs provided by Slurm and NVIDIA's DCGM are high-dimensional and exhibit complex patterns, as our data analysis in  Section~\ref{sec4} demonstrates. Given the nature of our prediction tasks, our framework design includes a regression model for predicting resource utilization values before job submission, and a classification model for runtime power prediction. The first stage predictions offer more flexibility compared to those of the second stage, as the GPU node resources have yet to be assigned. Therefore, we use a regression model to predict the continuous values of resource utilization and power usage.

In contrast, the second stage of our framework focuses on predicting the average runtime power consumption and is formulated as a classification problem. As we show in Section~\ref{sec4}, the average power consumption of 32,322 VASP job samples varies widely. Since our goal is to improve GPU power utilization efficiency through power capping using \textbf{runtime} predictions with a \textbf{low-overhead} prediction framework, fine-grained estimations using regression models become impractical, as they necessitate adjusting the power capping values frequently. Instead, as Figure~\ref{fig:overview-framework} presents, we create four different classes for the average power utilization (in watts) after observing the power usage distribution plot in Figure~\ref{fig:res-util-dist}. This classification motivates the use of our framework at runtime, where power capping values are adjusted only when the predicted target class changes. We further detail this implementation in Section~\ref{sec6.2}.

\paragraph{Data Preparation} As Figure~\ref{fig:overview-framework} demonstrates, we first divide the data into training and testing sets using an 80 to 20\% ratio before model training to preserve the time ordering of GPU jobs. Next, we fit an ordinal encoder from the \texttt{scikit-learn} library \cite{scikit} on the categorical Slurm features (user, job, account name, and scientific category) in the training set and apply the fitted encoder to the test set. The encoder assigns integer values to observed categories and maps unseen test categories to -1, making it particularly suitable for deployment scenarios where new categorical values (e.g., previously unseen job names) may occur. The final step of data preparation is scaling the training and test datasets using a standard scaler from the \texttt{scikit-learn} library \cite{scikit}, which standardizes features to a zero mean and unit variance. This is a common best practice for handling values with differing scales, as seen in the VASP GPU jobs.

\paragraph{ML Model Selection} For the first stage predictions, we evaluate several regression models, including tree-based ensemble methods such as XGBoost \cite{xgboost, xgboost_docs}, LightGBM \cite{lightgbm_paper, lightgbm_docs}, and a simple neural network. We apply hyperparameter tuning to all models. Overall, XGBoost and LightGBM achieve high prediction accuracy with low computational overhead, making them suitable for our prediction framework. In contrast, the neural network model requires significantly longer training times while offering lower accuracy. Based on these observations, we adopt the LightGBM regression model for the first stage predictions. For the second-stage runtime average power prediction task, we similarly compare multiple models and select the LightGBM classifier due to its high accuracy and low training and inference time overhead.

\paragraph{Evaluation Metrics} 
To evaluate our first-stage predictions, we use traditional metrics such as the mean absolute error (MAE), symmetric accuracy ($\mathrm{Acc}_{\text{sym}}$), and the coefficient of determination ($R^2$). MAE reports the average magnitude of numerical errors in the regression model predictions. $\mathrm{Acc}_{\text{sym}}$ tells us how closely the predictions match the actual utilization values, and the $R^2$ score measures how well the model predictions fit the actual data points.

\begin{equation*}
\mathrm{MAE} = \frac{1}{N} \sum_{i=1}^{N} \left| y_i - \hat{y}_i \right|
\end{equation*}

\begin{equation*}
\mathrm{Acc}_{\text{sym}} = \frac{1}{N} \sum_{i=1}^{N} 
\min\left( \frac{\hat{y}_i}{y_i}, \frac{y_i}{\hat{y}_i} \right)
\end{equation*}

\begin{equation*}
R^2 = 1 - \frac{\sum_{i=1}^{N} (y_i - \hat{y}_i)^2}
{\sum_{i=1}^{N} (y_i - \bar{y})^2}
\end{equation*}

Here, $y_i$ and $\hat{y}_i$ denote the true and predicted values for job $i$, and $N$ is the number of data points in the test set. For the second stage, where we predict the average runtime power of GPU jobs, we use the accuracy and macro average F1 score metrics. Accuracy measures the ratio of correctly classified test cases, while the F1 score reflects the model's prediction performance across all classes equally. In the following, $TP_c$, $FP_c$, and $FN_c$ denote the number of true positives, false positives, and false negatives for class $c$, respectively.

\begin{equation*}
\mathrm{Accuracy} = \frac{\text{Number of correct predictions}}{\text{Total number of predictions}}
\end{equation*}

\begin{equation*}
\text{F1 score} = \frac{1}{C} \sum_{c=1}^{C} \frac{2\,TP_c}{2\,TP_c + FP_c + FN_c}
\end{equation*}

\section{Experimental Results}\label{sec6}
In this section, we present the experimental results of our two-stage framework, differentiating the stages as methods for predicting resources before and during job execution. 

\subsection{Before Job Execution: Prediction with Slurm Submission Features}

Our first strategy to estimate the maximum GPU and memory utilization percentages, as well as the average power consumption of VASP jobs, is to use Slurm features that are available before the execution of the corresponding GPU job. Predicting the resource utilization of a job based solely on its Slurm submission features allows users to estimate resource consumption before their jobs execute. With this knowledge, they can scale their resource needs appropriately and avoid over- or under-provisioning GPU nodes or power usage. 

Table~\ref{tab:vasp_results} summarizes the first-stage prediction results for VASP jobs across all three target variables using only six training features. The framework achieves strong symmetric accuracy for all targets, with maximum GPU utilization performing best at 0.97, followed by average power at 0.94 and GPU memory utilization at 0.88. The $R^2$ scores have a complementary trend: while the model captures 59\% of the variance in maximum GPU utilization, it explains progressively more variance in GPU memory utilization and average power, reaching 0.70 and 0.84, respectively. These results suggest that these latter targets are more predictable from the available features.

\begin{table}[b]
  \centering
  \caption{First-Stage Prediction Performance for VASP Jobs}
  \label{tab:vasp_results}
  \rowcolors{2}{white}{gray!10}
  \begin{tabular}{l|
                  >{\centering\arraybackslash}p{1.8cm}
                  >{\centering\arraybackslash}p{1.2cm}
                  >{\centering\arraybackslash}p{1.0cm}}
    \toprule
    \rowcolor{white}
    \textbf{Target Variable} & \textbf{Sym. Acc} ($\uparrow$) & \textbf{MAE} ($\downarrow$) & \textbf{$R^2$} ($\uparrow$) \\
    \midrule
    Max. GPU Utilization   & 0.97 & 2.39\% & 0.59 \\
    Max. Memory Utilization & 0.88 & 2.58\% & 0.70 \\
    Average Power      & 0.94 & 8.57W & 0.84 \\
    \bottomrule
  \end{tabular}
\end{table}

\begin{figure*}[!b]
\centering
    \includegraphics[width=0.8\textwidth]{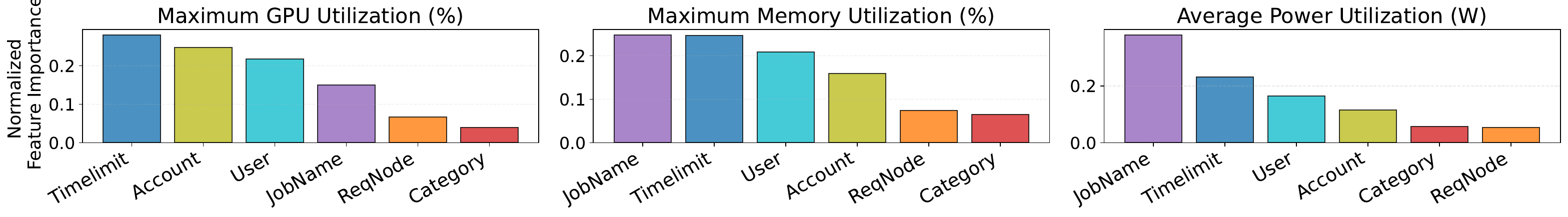}
    \caption{Before job execution prediction: Normalized feature importance scores from the LightGBM regression models for each target variable.}
    \label{fig:vasp-res}
\end{figure*} 

To compare our prediction framework with recent work, we implement the UoPC \cite{antici25isc} power prediction model as a baseline method. UoPC employs the $k$-nearest neighbors (KNN) \cite{fix1985discriminatory} model, trained separately for each user using Slurm submission parameters to predict the average and maximum power consumption of jobs on the Fugaku supercomputer. We first observe that UoPC's user-specific model training strategy does not apply to users with a small number of VASP job submissions.  In our dataset, this results in the exclusion of 59 out of 128 VASP users from model training. As a result, a direct one-to-one comparison is inherently limited, since our framework makes predictions for all users, regardless of the number of jobs submitted, whereas UoPC does not generalize beyond frequent users. Even when restricting the baseline comparison to users with sufficient training data, our LightGBM-based framework consistently outperforms UoPC across all target variables. Specifically, UoPC yields negative $R^2$ values for GPU utilization, memory utilization, and average power prediction, indicating poor generalization to unseen jobs while also exhibiting higher MAE (9.5\%, 5.2\%, and 10.3 watts, respectively) and lower accuracy (0.89, 0.75, and 0.93, respectively).

Lastly, in Figure~\ref{fig:vasp-res}, we present the input feature importance results derived from the LightGBM models for each target prediction variable. In general, the time limit, job name, and username are the most important training features across different target variables. The consistent importance of job names shows that users encode certain characteristics of their VASP workloads in their batch job names. The time limit further provides an estimate of the expected batch job intensity and workload scale, which correlates with resource and power demands. Our observations align with prior work \cite{antici25isc}, which states that using metadata (username and job names) and requested resources as training features improves predictions of power and resource utilization in HPC systems. 

The results in this subsection show that our first-stage framework accurately captures GPU resource and average power utilization behavior from a small set of Slurm submission features before job execution, with an accuracy of up to 97\%. Our framework is therefore suitable for practical deployment in scheduling and resource management on large-scale heterogeneous HPC systems. Next, we focus on predicting the average power consumption of VASP GPU workloads while they are running, using runtime DCGM metrics. 

\subsection{During Job Execution: Time Series Prediction with DCGM Runtime Metrics}\label{sec6.2} 
For VASP jobs, resource utilization mainly depends on the input size \cite{zhao2024sc}. While access to VASP job inputs is limited to the corresponding user, runtime DCGM metrics can help estimate the input size of VASP jobs and result in more accurate predictions. Furthermore, we can use the runtime metrics to enable the dynamic management of GPU node resources. With these motivations, we include the time series DCGM metrics in Table~\ref{tab:dcgm_metrics}, in addition to the job submission parameters provided by the Slurm workload manager, to predict the average power consumption of any given VASP GPU job during its execution. While our results show the predictions for VASP GPU jobs only, this framework can be applied to other GPU jobs for average power prediction as well. We elaborate further on the generalizability of our framework to additional GPU workloads in Section~\ref{disc}. To the best of our knowledge, this work is the first attempt to predict the average power usage of GPU jobs at runtime using time series GPU monitoring metrics, ultimately aiming to dynamically manage the power utilization of heterogeneous HPC applications.

\begin{figure*}[t]\centering
    \includegraphics[width=0.7\textwidth]{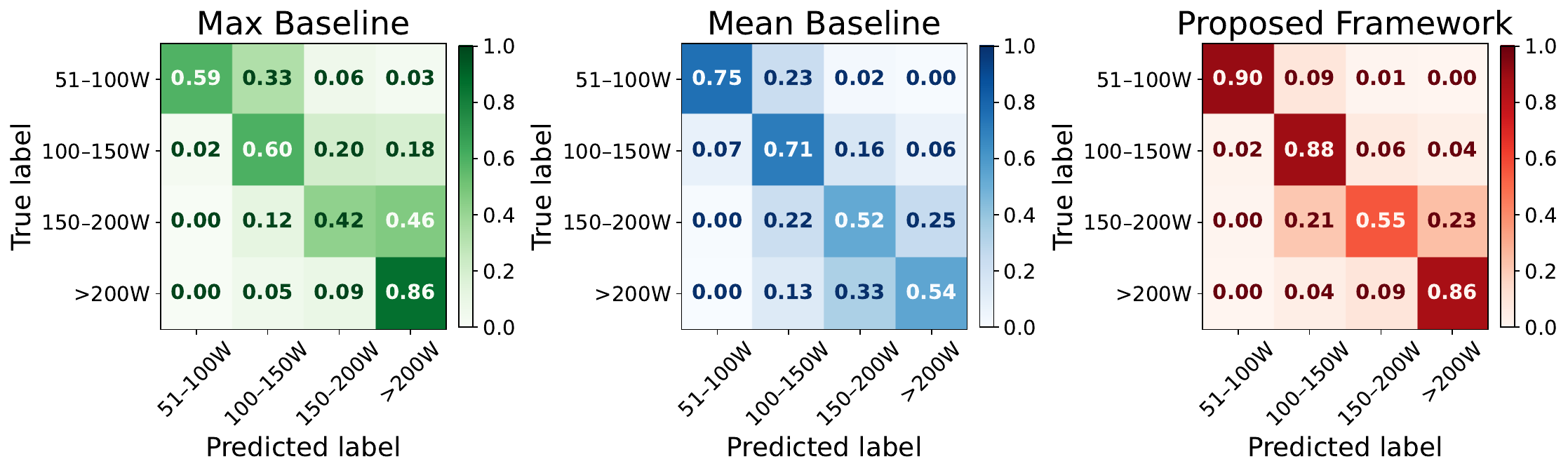}
    \caption{During job execution power prediction: Normalized confusion matrices comparing baseline methods and the proposed framework.} 
    \label{fig:cm2}
\end{figure*}

The data we use in this set of experiments are time series, as explained in Section~\ref{sec3}, and we apply padding to the static input parameters obtained from Slurm to mimic this time series behavior. We employ a sliding window approach, using a training window size of three to predict the average power consumption value at the next time step. The choice of a window size of three is motivated by the fact that scheduling decisions are made at 30-second intervals at Perlmutter \cite{nersc_jobs_best_practices}, and the DCGM metrics are collected every 10 seconds during a GPU job's execution. At those 10-second intervals, Perlmutter GPU nodes report four distinct measurements for each DCGM metric from four different A100 GPUs. Note that Perlmutter’s power capping feature enforces the same power limit across all GPUs assigned to a job \cite{perlmutter_power_cap}. To support dynamic power management with power capping strategies using our time series–based power prediction framework, we use the mean values of the DCGM metrics reported across all nodes and GPUs at each time step. These averaged DCGM metrics are then used to predict the average power consumption for a given GPU job at the next time step. As a result, for each prediction instance at time $t$ in the test dataset, we predict the average power consumption of a VASP job at time $t+1$ (aggregated across all GPUs and nodes) using the DCGM metrics from times $t$, $t-1$, and $t-2$. When adopted by other systems, our choice of window size may change based on the requirements.

\begin{table}[b]
  \centering
  \caption{Runtime Power Prediction Performance for VASP Jobs}
  \label{tab:stage-2-perf-metrics}
  \rowcolors{2}{white}{gray!10}
  \begin{tabular}{l|
                  >{\centering\arraybackslash}p{1.2cm}
                  >{\centering\arraybackslash}p{1.6cm}}
    \toprule
    \rowcolor{white}
    \textbf{Model} & \textbf{Acc} ($\uparrow$) & \textbf{F1 Score} ($\uparrow$) \\
    \midrule
    Max Baseline        & 0.63 & 0.62 \\
    Mean Baseline       & 0.63 & 0.62 \\
    \textbf{Proposed Framework} & \textbf{0.82} & \textbf{0.80} \\
    \bottomrule
  \end{tabular}
\end{table}

\begin{figure}[t]
    \includegraphics[width=\linewidth]{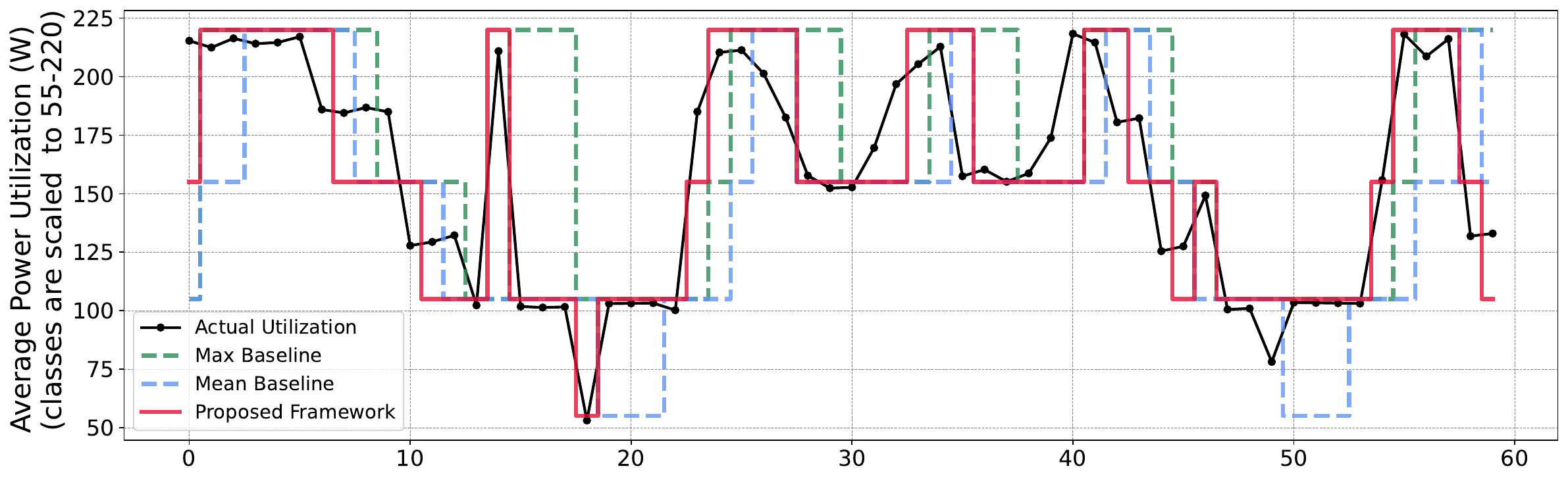}
    \caption{Test samples from a 10-minute VASP job execution window, showing runtime predictions of average power consumption class.} 
    \label{fig:stage2-ex}
\end{figure}

\begin{figure}[b]
    \includegraphics[width=\linewidth]{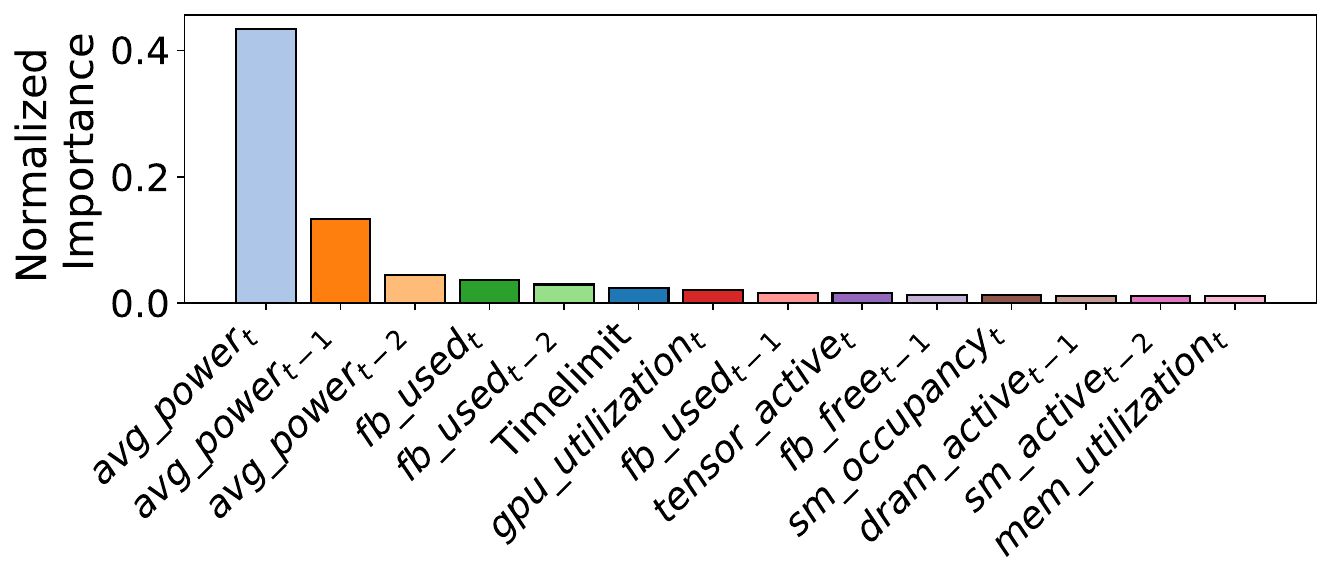}
    \caption{During job execution prediction: Normalized feature importance scores from the LightGBM classifier.} 
    \label{fig:feat2}
\end{figure} 

Using the methodology from Section~\ref{sec5}, we apply the second stage of our framework to predict the average power usage of VASP jobs during their execution. We compare our results with two simple baseline methods, namely the mean and maximum values of the average power usage from times $t$, $t-1$, and $t-2$, to illustrate why an ML model is preferable for this prediction task. Our framework substantially outperforms both baselines, improving accuracy from 0.63 to 0.82 and the F1 score from 0.62 to 0.80. The normalized confusion matrices in Figure~\ref{fig:cm2} further highlight the performance of ML-based predictions. The results from naive prediction methods are as expected, with their most accurate predictions corresponding to the highest and lowest utilization classes for the maximum and mean baselines, respectively. Our ML-based prediction strategy, on the other hand, accurately estimates the average power usage class for the subsequent time step in the majority of cases, resulting in more stable and reliable runtime predictions. We also observe that our framework's misclassifications occur between neighboring utilization classes rather than abrupt changes. Furthermore, we illustrate the ML-based predictions in comparison with naive baselines in Figure~\ref{fig:stage2-ex} using 60 sample data points (corresponding to a 10-minute run) from the VASP dataset. The average power utilization class at time $t+1$ is best tracked with the ML-based approach, whereas the naive baselines tend to overpredict or underpredict.

Lastly, we show the normalized input feature importances in Figure~\ref{fig:feat2} from the time series-based LightGBM classifier model, where we observe that the DCGM metrics demonstrate greater importance than the Slurm input features. The average power consumption value at time $t$ is the most important input parameter, followed by the DCGM metrics related to memory and SM utilization. These results support our initial claim that DCGM GPU performance metrics, augmented with Slurm submission features, can help accurately predict the runtime power consumption of GPU jobs.

\section{Discussion}\label{disc}
In this work, we provide a case study on the resource and power usage of VASP applications and present a generalizable prediction framework for heterogeneous HPC applications. We aim to motivate the use of machine learning techniques to predict GPU workloads' resource consumption for better scheduling performance before they start running on the system. Therefore, in addition to VASP, we conduct experiments with 5 other applications from Perlmutter to support the generalizability of our framework. These applications are LAMMPS \cite{thompson2022lammps}, Espresso \cite{giannozzi2020quantum}, Atlas \cite{ATLASSoftwareDocs}, and E3SM \cite{taylor2023simple}, for which Slurm and DCGM metrics from March 2025 with 134, 363, 655, and 267 GPU jobs available, respectively. 

For the first stage predictions using Slurm metrics only, we observe similar trends to those for VASP, with the average power prediction accuracy exceeding 0.73 for E3SM and approaching 0.88 for LAMMPS, Espresso, and Atlas. Similarly, the maximum GPU utilization predictions result in accuracy ranging from 0.69 to 0.80. In contrast, with this small set of application data, memory utilization is more challenging to predict, with an accuracy ranging from 0.49 to 0.69.

The second-stage runtime average power predictions also demonstrate high accuracy. For LAMMPS, Espresso, Atlas, and E3SM, the accuracy reaches 0.85, 0.81, 0.92, and 0.85, respectively. Our framework improves F1 scores over both maximum and mean naive baselines, ranging from 0.69 to 0.86. These results further support our claim that the proposed prediction framework, which we designed based on a detailed analysis of VASP workloads, achieves high accuracy in resource and power usage prediction across diverse GPU workloads and generalizes to other GPU-accelerated jobs in production HPC systems.

\section{Conclusion and Future Work}\label{sec8}
In this work, we provide an application-level resource utilization analysis of VASP jobs running on Perlmutter's GPU nodes. By identifying the mismatch between GPU and GPU memory utilization, as well as the wide range of power usage behaviors, we highlight the need for resource prediction mechanisms in heterogeneous computing platforms to enable more efficient use of GPU nodes and available power. To this end, we develop a prediction method that can inform Perlmutter users about their job's estimated resource consumption at the time of job submission. Using Slurm accounting data from 32,322 different VASP GPU jobs, our first-stage predictions reach an accuracy of at least 88\%. Additionally, by using time series DCGM metrics, we provide a novel approach for predicting the average power consumption of unseen time series data every 10 seconds and achieve up to 92\% accuracy. Our work demonstrates the effectiveness of DCGM metrics in capturing application characteristics that are relevant to power consumption. 

While this work focuses on VASP GPU jobs running on Perlmutter, our approach is generalizable across diverse GPU workloads (demonstrated on LAMMPS, Espresso, Atlas, and E3SM) and other heterogeneous HPC systems without requiring any modifications to the proposed methodology. We present our framework as a reusable tool for the broader HPC community, and we support reproducibility by open-sourcing our resource utilization analysis notebooks and prediction framework. As part of our future work, we plan to apply the first stage of our prediction methodology to Perlmutter GPU jobs before scheduling and to make informed decisions about GPU resources in order to improve energy-efficient scheduling efforts and increase overall system throughput. Moreover, we aim to extend our time series-based power prediction model using DCGM runtime metrics by deploying it for real-time inference for GPU applications and complementing it with power capping strategies. Our goal is to enable dynamic power management and enhance the energy efficiency of HPC systems, even under strict power constraints. 

\section*{Acknowledgment}

This research was supported by the National Energy Research Scientific Computing Center (NERSC), a U.S. Department of Energy Office of Science User Facility operated under Contract No. DE-AC02-05CH11231.

\bibliographystyle{IEEEtran}
\bibliography{references}

@INPROCEEDINGS{nichols2022resource,
  author={Nichols, Daniel and Marathe, Aniruddha and Shoga, Kathleen and Gamblin, Todd and Bhatele, Abhinav},
  booktitle={2022 IEEE International Parallel and Distributed Processing Symposium (IPDPS)}, 
  title={Resource Utilization Aware Job Scheduling to Mitigate Performance Variability}, 
  year={2022},
  volume={},
  number={},
  pages={335-345},
  doi={10.1109/IPDPS53621.2022.00040}}

@INPROCEEDINGS{maloney2024analyzing,
  author={Maloney, Samuel and others},
  booktitle={2024 IEEE 36th International Symposium on Computer Architecture and High Performance Computing (SBAC-PAD)}, 
  title={Analyzing HPC Monitoring Data With a View Towards Efficient Resource Utilization}, 
  year={2024},
  volume={},
  number={},
  pages={170-181},
  doi={10.1109/SBAC-PAD63648.2024.00023}}

@inproceedings{tanash2021ensemble,
author = {Tanash, Mohammed and Yang, Huichen and Andresen, Daniel and Hsu, William},
title = {Ensemble Prediction of Job Resources to Improve System Performance for Slurm-Based HPC Systems},
year = {2021},
isbn = {9781450382922},
publisher = {Association for Computing Machinery},
address = {New York, NY, USA},
url = {https://doi.org/10.1145/3437359.3465574},
doi = {10.1145/3437359.3465574},
booktitle = {Practice and Experience in Advanced Research Computing 2021: Evolution Across All Dimensions},
articleno = {21},
numpages = {8},
series = {PEARC '21}
}

@INPROCEEDINGS{thonglek2019improving,
  author={Thonglek, Kundjanasith and others},
  booktitle={2019 IEEE International Conference on Cluster Computing (CLUSTER)}, 
  title={Improving Resource Utilization in Data Centers using an LSTM-based Prediction Model}, 
  year={2019},
  volume={},
  number={},
  pages={1-8},
  doi={10.1109/CLUSTER.2019.8891022}}

@inproceedings{sencan2025gpu,
author = {Sencan, Efe and Kulkarni, Dhruva and Coskun, Ayse and Konate, Kadidia},
title = {Analyzing GPU Utilization in HPC Workloads: Insights from Large-Scale Systems},
year = {2025},
isbn = {9798400713989},
publisher = {Association for Computing Machinery},
address = {New York, NY, USA},
url = {https://doi.org/10.1145/3708035.3736010},
doi = {10.1145/3708035.3736010},
booktitle = {Practice and Experience in Advanced Research Computing 2025: The Power of Collaboration},
articleno = {14},
numpages = {8},
location = {
},
series = {PEARC '25}
}

@inproceedings{li2023analyzing,
  title={Analyzing resource utilization in an hpc system: A case study of nersc’s perlmutter},
  author={Li, Jie and Michelogiannakis, George and Cook, Brandon and Cooray, Dulanya and Chen, Yong},
  booktitle={International Conference on High Performance Computing},
  pages={297--316},
  year={2023},
  organization={Springer}
}

@inproceedings{menear2025classification,
author = {Menear, Kevin and Acosta, Joe and Duplyakin, Dmitry},
title = {Classification of HPC Job Power Consumption Using a Rich Feature Set},
year = {2025},
isbn = {9798400713989},
publisher = {Association for Computing Machinery},
address = {New York, NY, USA},
url = {https://doi.org/10.1145/3708035.3736084},
doi = {10.1145/3708035.3736084},
booktitle = {Practice and Experience in Advanced Research Computing 2025: The Power of Collaboration},
articleno = {65},
numpages = {6},
location = {
},
series = {PEARC '25}
}

@book{fix1985discriminatory,
  title={Discriminatory analysis: nonparametric discrimination, consistency properties},
  author={Fix, Evelyn},
  volume={1},
  year={1985},
  publisher={USAF school of Aviation Medicine}
}

@article{hafner2008ab,
  title={Ab-initio simulations of materials using VASP: Density-functional theory and beyond},
  author={Hafner, J{\"u}rgen},
  journal={Journal of computational chemistry},
  volume={29},
  number={13},
  pages={2044--2078},
  year={2008},
  publisher={Wiley Online Library}
}

@INPROCEEDINGS{agelastos2014lightweight,
  author={Agelastos, Anthony and Allan, Benjamin and Brandt, Jim and Cassella, Paul and Enos, Jeremy and Fullop, Joshi and Gentile, Ann and Monk, Steve and others},
  booktitle={SC '14: Proceedings of the International Conference for High Performance Computing, Networking, Storage and Analysis}, 
  title={The Lightweight Distributed Metric Service: A Scalable Infrastructure for Continuous Monitoring of Large Scale Computing Systems and Applications}, 
  year={2014},
  volume={},
  number={},
  pages={154-165},
  doi={10.1109/SC.2014.18}}

@article{scikit,
  title={Scikit-learn: {M}achine {L}earning in {P}ython},
  author={Pedregosa, Fabian and Varoquaux, Ga{\"e}l and Gramfort, Alexandre and Michel, Vincent and Thirion, Bertrand and Grisel, Olivier and Blondel, Mathieu and others},
  journal={the Journal of Machine Learning Research},
  volume={12},
  pages={2825--2830},
  year={2011},
  publisher={JMLR. org}
}

@INPROCEEDINGS{antici25isc,
  author={Antici, Francesco and Borghesi, Andrea and Domke, Jens and Kiziltan, Zeynep},
  booktitle={ISC High Performance 2025 Research Paper Proceedings (40th International Conference)}, 
  title={UoPC: A User-Based Online Framework to Predict Job Power Consumption in HPC Systems}, 
  year={2025},
  volume={},
  number={},
  pages={1-12},
  doi={}}

@INPROCEEDINGS{zhao2024sc,
  author={Zhao, Zhengji and Austin, Brian and Rrapaj, Ermal and Wright, Nicholas J.},
  booktitle={SC24-W: Workshops of the International Conference for High Performance Computing, Networking, Storage and Analysis}, 
  title={Understanding VASP Power Profiles on NVIDIA A100 GPUs}, 
  year={2024},
  volume={},
  number={},
  pages={1496-1505},
  doi={10.1109/SCW63240.2024.00189}}

@inproceedings{lightgbm_paper,
author = {Ke, Guolin and others},
title = {LightGBM: a highly efficient gradient boosting decision tree},
year = {2017},
isbn = {9781510860964},
publisher = {Curran Associates Inc.},
address = {Red Hook, NY, USA},
booktitle = {Proceedings of the 31st International Conference on Neural Information Processing Systems},
pages = {3149–3157},
numpages = {9},
location = {Long Beach, California, USA},
series = {NIPS'17}
}

@inproceedings{xgboost,
author = {Chen, Tianqi and Guestrin, Carlos},
title = {XGBoost: A Scalable Tree Boosting System},
year = {2016},
isbn = {9781450342322},
publisher = {Association for Computing Machinery},
address = {New York, NY, USA},
url = {https://doi.org/10.1145/2939672.2939785},
doi = {10.1145/2939672.2939785},
booktitle = {Proceedings of the 22nd ACM SIGKDD International Conference on Knowledge Discovery and Data Mining},
pages = {785–794},
numpages = {10},
location = {San Francisco, California, USA},
series = {KDD '16}
}

@INPROCEEDINGS{hpca2015gpubuild,
  author={Wu, Gene and others},
  booktitle={2015 IEEE 21st International Symposium on High Performance Computer Architecture (HPCA)}, 
  title={GPGPU performance and power estimation using machine learning}, 
  year={2015},
  volume={},
  number={},
  pages={564-576},
  doi={10.1109/HPCA.2015.7056063}}

@inproceedings{tardis2025power,
  title={Power-aware scheduling for multi-center hpc electricity cost optimization},
  author={Hossain, Abrar and Abdurahman, Abubeker and Islam, Mohammad A and Ahmed, Kishwar},
  booktitle={Workshop on Job Scheduling Strategies for Parallel Processing},
  pages={1--20},
  year={2025},
  organization={Springer}
}

@inproceedings{onlineincremental2023,
author = {Tian, Xuesen and others},
title = {An Online Incremental Learning Framework for HPC Job Power Consumption Prediction},
year = {2023},
isbn = {9781450399883},
publisher = {Association for Computing Machinery},
address = {New York, NY, USA},
url = {https://doi.org/10.1145/3606043.3606068},
doi = {10.1145/3606043.3606068},
booktitle = {Proceedings of the 2023 7th International Conference on High Performance Compilation, Computing and Communications},
pages = {176–183},
numpages = {8},
location = {Jinan, China},
series = {HP3C '23}
}

@inproceedings{framework2024performance,
  title={A framework and methodology for performance prediction of hpc workloads},
  author={Orteu, J{\'u}lia and Clasc{\`a}, Marc and Labarta, Jes{\'u}s and Jennings, Elise and Andersson, Stefan and Garcia-Gasulla, Marta},
  booktitle={International Conference on Parallel Processing and Applied Mathematics},
  pages={114--127},
  year={2024},
  organization={Springer}
}

@INPROCEEDINGS{summit2020temporal,
  author={Li, Chengcheng and others},
  booktitle={2021 IEEE International Conference on Cluster Computing (CLUSTER)}, 
  title={The Challenge of Disproportionate Importance of Temporal Features in Predicting HPC Power Consumption}, 
  year={2021},
  volume={},
  number={},
  pages={632-636},
  doi={10.1109/Cluster48925.2021.00094}}

@article{polaris2025energy,
  title={Extracting Practical, Actionable Energy Insights from Supercomputer Telemetry and Logs},
  author={Cornelius, Melanie and Cross, Greg and Shilpika, Shilpika and Dearing, Matthew T and Lan, Zhiling},
  journal={arXiv preprint arXiv:2505.14796},
  year={2025}
}

@article{karimi2024profiling,
  title={Profiling and modeling of power characteristics of leadership-scale hpc system workloads},
  author={Karimi, Ahmad Maroof and Sattar, Naw Safrin and Shin, Woong and Wang, Feiyi},
  journal={arXiv preprint arXiv:2402.00729},
  year={2024}
}

@inproceedings{taylor2023simple,
author = {Taylor, Mark and others},
title = {The Simple Cloud-Resolving E3SM Atmosphere Model Running on the Frontier Exascale System},
year = {2023},
isbn = {9798400701092},
publisher = {Association for Computing Machinery},
address = {New York, NY, USA},
url = {https://doi.org/10.1145/3581784.3627044},
doi = {10.1145/3581784.3627044},
booktitle = {Proceedings of the International Conference for High Performance Computing, Networking, Storage and Analysis},
articleno = {7},
numpages = {11},
location = {Denver, CO, USA},
series = {SC '23}
}

@article{giannozzi2020quantum,
  title={Quantum ESPRESSO toward the exascale},
  author={Giannozzi, Paolo and Baseggio, Oscar and Bonf{\`a}, Pietro and Brunato, Davide and Car, Roberto and Carnimeo, Ivan and Cavazzoni, Carlo and De Gironcoli, Stefano and others},
  journal={The Journal of chemical physics},
  volume={152},
  number={15},
  year={2020},
  publisher={AIP Publishing}
}

@misc{ATLASSoftwareDocs,
  author       = {{Athena Developers}},
  title        = {ATLAS Software Documentation},
  howpublished = {\url{https://atlas-software.docs.cern.ch}},
  note         = {Structured tutorials and guides for software used in the ATLAS experiment at CERN},
  year         = {2025},
  url          = {https://atlas-software.docs.cern.ch}
}

@article{thompson2022lammps,
  title={LAMMPS-a flexible simulation tool for particle-based materials modeling at the atomic, meso, and continuum scales},
  author={Thompson, Aidan P and Aktulga, H Metin and Berger, Richard and Bolintineanu, Dan S and Brown, W Michael and Crozier, Paul S and In't Veld, Pieter J and Kohlmeyer, Axel and Moore, Stan G and Nguyen, Trung Dac and others},
  journal={Computer physics communications},
  volume={271},
  pages={108171},
  year={2022},
  publisher={Elsevier}
}

@misc{xgboost_docs,
  author       = {{XGBoost Developers}},
  title        = {XGBoost Documentation},
  year         = {2025},
  howpublished = {Retrieved from \url{https://xgboost.readthedocs.io/en/stable/}},
}

@misc{lightgbm_docs,
  author       = {{Microsoft/LightGBM Developers}},
  title        = {LightGBM Documentation},
  howpublished = {Retrieved from \url{https://lightgbm.readthedocs.io/en/latest/index.html}},
  year={2026}
}

@misc{slurmurl,
  author       = {{SchedMD}},
  title        = {Slurm Workload Manager Documentation},
  howpublished = {Retrieved from \url{https://slurm.schedmd.com/documentation.html}},
  year={2026}
}

@misc{vasp,
  author    = {{VASP Developers}},
  title        = {{Vienna Ab initio Simulation Package (VASP)}},
  howpublished = {Retrieved from \url{https://www.vasp.at/}},
  year         = {2026}
}

@misc{nersc_jobs_best_practices,
  title        = {{Best Practices for Running Jobs at NERSC}},
  howpublished = {Retrieved from \url{https://docs.nersc.gov/jobs/best-practices/}},
  year         = {2025},
  author       = {{National Energy Research Scientific Computing Center (NERSC)}}
}

@misc{perlmutter_docs,
    title        = {{Perlmutter Architecture Documentation}},
    author       = {{National Energy Research Scientific Computing Center (NERSC)}},
    year={n.d},
    howpublished = {Retrieved from \url{https://docs.nersc.gov/systems/perlmutter/architecture/}},
}

@misc{perlmutter_power_cap,
  title        = {{GPU Power Capping on Perlmutter}},
  howpublished = {Retrieved from \url{https://docs.nersc.gov/jobs/power-capping/}},
  author       = {{National Energy Research Scientific Computing Center (NERSC)}},
  year={n.d}
}

@misc{nvidia_dcgm_docs,
  title        = {NVIDIA Data Center GPU Manager (DCGM) Documentation},
  author       = {{NVIDIA Corporation}},
  year         = {2026},
  howpublished = {Retrieved from \url{https://docs.nvidia.com/datacenter/dcgm/latest/index.html}}
}

\end{document}